\documentclass[prd, aps, superscriptaddress, preprintnumbers, twocolumn, floatfix, nofootinbib]{revtex4}
\pdfoutput=1

\usepackage{amsfonts}
\usepackage{amsmath}
\usepackage{amssymb}
\usepackage{bm}
\usepackage{dcolumn}
\usepackage{graphicx}   
\usepackage[latin1]{inputenc}
\usepackage{latexsym}
\usepackage{rotating}
\usepackage{hyperref}
\usepackage{graphicx}
\usepackage{color}
\usepackage{cases}

\newcommand\be{\begin{equation}}
\newcommand\ba{\begin{eqnarray}}
\newcommand\ee{\end{equation}}
\newcommand\ea{\end{eqnarray}}

\newcommand{\Msol}{\ensuremath{M_{\odot}}}
\newcommand{\nn}{\nonumber \\}
\newcommand{\gsim}{\mathrel{\hbox{\rlap{\lower.55ex \hbox {$\sim$}}
                   \kern-.3em \raise.4ex \hbox{$>$}}}}
\newcommand{\lsim}{\mathrel{\hbox{\rlap{\lower.55ex \hbox {$\sim$}}
                   \kern-.3em \raise.4ex \hbox{$<$}}}}

\begin{document}

\title {Early Structure Formation from Cosmic String Loops \\in Light of Early JWST Observations}

\author{Hao Jiao}
\email{hao.jiao@mail.mcgill.ca}
\affiliation{Department of Physics, McGill University, Montr\'{e}al, QC, H3A 2T8, Canada} 

\author{Robert Brandenberger}
\email{rhb@physics.mcgill.ca}
\affiliation{Department of Physics, McGill University, Montr\'{e}al, QC, H3A 2T8, Canada} 

\author{Alexandre Refregier}
\email{alexandre.refregier@phys.ethz.ch}
\affiliation{Institute for Particle Physics and Astrophysics, ETH, Wolfgang-Pauli-Strasse 27, CH-8093 Zurich, Switzerland}

\date{\today}

%%%%%%%%%%%%%%%%%%%%%%%%%%%%%%%%%%%%%%%%%%%%%%%%%%%%%%%%%%%%%%%%%%%%%%%%%%%%%%%%%%%%%%%%%%%%%%

\begin{abstract}

Cosmic strings, if they exist,  source nonlinear and non-Gaussian perturbations all the way back to the time of equal matter and radiation (and earlier).  Here, we compute the mass function of halos seeded by a scaling distribution of cosmic string loops, and we compare the results with the predictions of the standard Gaussian $\Lambda$CDM model.  Assuming a simple linear relation between stellar mass and halo mass, we also compute the stellar mass function. The contribution of cosmic strings dominates at sufficiently high redshifts $z > z_c$ where $z_c$ depends on the mass of the halo and on the mass per unit length $\mu$ of the strings and is of the order $z_c \sim 12$ for $G\mu = 10^{-8}$. We find that strings with this value of $G\mu$ can explain the preliminary JWST data on the high redshift stellar mass density.  Based on an extreme value statistic, we find that the mass of the heaviest expected string-seeded galaxy for the current JWST sky coverage is compatible with the heaviest detected galaxy.  Given the uncertainties in the interpretation of the JWST data, we discuss predictions for higher redshift observations. 

\end{abstract}
%%%%%%%%%%%%%%%%%%%%%%%%%%%%%%%%%%%%%%%%%%%%%%%%%%%%%%%%%%%%%%%%%%%%%%%%%%%%%%%%%%%%%%%%%%%%%%

\maketitle

%%%%%%%%%%%%%%%%%%%%%%%%%%%%%%%%%%%%%%%%%%%%%%%%%%%%%%%%%%%%%%%%%%%%%%%%%%%%%%%%%%%%%%%%%%%%%%
%%%%%%%%%%%%%%%%%%%%%%%%%%%%%%%%%%%%%%%%%%%%%%%%%%%%%%%%%%%%%%%%%%%%%%%%%%%%%%%%%%%%%%%%%%%%%%
\section{Introduction} 
\label{sec:intro}

New windows to probe the early Universe are opening up. For example, the James Webb Space Telescope (JWST) is allowing us to better explore the early stages of galaxy formation. Preliminary data and its interpretation \cite{JWST-data} from the JWST telescope indicate that the mass fraction in early halos exceeds what is predicted by the current concordance model of early Universe cosmology, the $\Lambda$CDM model \cite{JWST-problems,JWST-EVS}, although the reader must be warned that there remain some systematic uncertainties concerning the interpretation of the data.  If the initial data and interpretation are confirmed, then a modification of the standard $\Lambda$CDM model will be required.

The standard $\Lambda$CDM model is based on the assumption that the primordial cosmological fluctuations are given by a nearly Gaussian random field with an almost scale-invariant spectrum, such as predicted by inflation \cite{Mukh} or by alternatives \cite{SGCflucts} such as String Gas Cosmology \cite{SGC} \footnote{See e.g. \cite{RHB-alt} for a review of some alternative scenarios.}.   In this paper, we will study the effects which a distribution of cosmic string loops may have had on early halo formation. Strings provide a highly non-Gaussian contribution to the density field, and hence could help with the generation of nonlinear halos at high redshifts \footnote{See also \cite{Biagetti} for a more general analysis of how primordial non-Gaussianities may help explain the JWST data.}.

There are other observational hints that the cosmological fluctuations in the standard $\Lambda$CDM model are insufficient to explain high-redshift observations.  For example, without postulating a prolonged phase of super-Eddington accretion it is not possible to explain the abundance (see e.g. \cite{Marta} for a review) of super-massive black holes at high redshifts \cite{Jerome}, and it was shown in \cite{Bryce1} that cosmic strings with values of $G\mu$ even lower than the gravitational radiation bound of \cite{CS-pulsar} provide a sufficient number of nonlinear seeds at high redshift to explain the data, and in \cite{Bryce2} it was shown, in addition, that in the case of superconducting cosmic string loops it is possible to satisfy the {\it direct collapse black hole criteria} at high enough redshift.

From the point of view of particle physics, it is well motivated to assume that there may have been a phase transition in the early universe which leads to the production of a network of cosmic strings in analogy to how a cooling transition in certain metals yields a network of line defects. Specifically, cosmic strings form in a symmetry breaking phase transition in any model beyond the Standard Model of particle physics in which the space of ground states has the topology of a circle (see e.g. \cite{CS-revs} for reviews of the role of cosmic strings and other topological defects in cosmology).
 
If nature is described by a particle physics model which admits cosmic string defects, then a network of strings will -- by causality arguments \cite{Kibble} -- form in the early universe and persist to the present time. The network consists of ``long'' strings (with curvature radius greater than the Hubble length) and a distribution of string loops.  The distribution scales in the sense that the statistical properties of the string network are independent of time if all lengths are scaled to the Hubble radius \footnote{We work in the context of a spatially flat Friedmann-Robertson-Walker-Lemaitre cosmology with scale factor $a(t)$,  $t$ being physical time,  and comoving coordinates ${\bf{x}}$. The Hubble radius is the inverse expansion rate,  and the cosmological redshift is denoted by $z(t)$.}.  Strings carry trapped energy which gravitates and hence contributes to structure formation. Specifically, strings lead to nonlinear density fluctuations at arbitrarily early times and can hence contribute to and even dominate early halo formation. In this paper, we will focus on the role of cosmic string loops (see e.g. \cite{CS-early} for early works on the role of cosmic strings in galaxy formation).

The distribution of cosmic strings is universal. In particular, it does not depend on the mass per unit length $\mu$ of the string, the one parameter which determines the gravitational effects of the strings. The value of $\mu$ is related to the energy scale $\eta$ when the string-forming phase transition takes place ($\mu \simeq \eta^2$).  The mass per unit length is usually described by the dimensionless parameter $G\mu$, where $G$ is Newton's gravitational constant \footnote{We are using natural units in which the speed of light, Planck's constant and Boltzmann's constant are all set to $1$. }.  The value of $G\mu$ is already constrained by cosmological observations: a value of $G\mu > 10^{-7}$ is ruled out from measurements of the angular power spectrum of the cosmic microwave background (CMB) \cite{CMBlimits}.  This bound is robust since it is independent of additional assumptions about the string loop distribution. Making assumptions about the loop distribution, bound as low as $G\mu < 10^{-10}$ can be derived from the upper limits on the spectrum of stochastic gravitational waves from pulsar timing array studies \cite{CS-pulsar}.

In this paper, we study the halo mass function induced by a scaling distribution of cosmic string loops, focusing in particular on the redshift dependence.  We find that the present JWST data on the abundance of stellar halo masses at redshifts of $z = 8$ and $z = 10$ could be explained for a value of $G\mu$ lower than the robust bound from CMB anisotropies.  The mass function decays only as a power law of $(1 + z)^{-1}$, not exponentially as it does at high redshifts in the $\Lambda$CDM model.  We also consider the {\it extreme value statistic} for the highest halo mass expected as a function of redshift in the cosmic string model.  We find that the largest and earliest observed halo masses at high redshifts can be explained with cosmic strings, making use of the same value of $G\mu$ as the one which fits the abundance of stellar halo mass.

In the following section, we briefly review the scaling distribution of cosmic string loops. This distribution is the basis for computing the stellar mass function of loop-seeded halos. This is discussed in Section 3. Section 4 focuses on the determination of the probability distribution of the expected largest mass halo by extreme value statistic analysis. We end with a summary of our findings and discussion of the results.

Our work is not the first to study early structure formation from strings.  Early reionization from string wakes (overdensities which form behind moving long strings \cite{wake}) was studied in \cite{Pogosian} (see also \cite{Duplessis}), and the contribution of string loops to reionization was analyzed in \cite{OlumVilenkin}.  A detailed analysis of early structure formation from string loops was presented in \cite{Shlaer}. Our analysis takes another look at this issue, focusing on the comparison with the preliminary JWST data and performing an {\it extreme value statistic analysis}. Note that cosmic strings also give rise to distinguished signatures in high redshift 21-cm maps \cite{Holder, Maibach}.

%%%%%%%%%%%%%%%%%%%%%%%%%%%%%%%%%%%%%%%%%%%%%%%%%%%%%%%%%%%%%%%%%%%%%%%%%%%%%%%%%%%%%%%%%%%%%%

\section{Review of the Cosmic String Scaling Solution} \label{review}

If nature is described by a particle physics model which admits cosmic string solutions, then a network of strings inevitably forms during the symmetry breaking phase transition in the early universe and persists to the present time \cite{Kibble}. The string network consists of ``infinite'' strings and string loops.  Here, ``infinite'' means strings with a curvature radius comparable to or larger than the Hubble radius $t$.  The long string network has a mean curvature radius $\xi$ which scales as $\xi \sim t$.  This implies that the statistical properties of the string network are independent of time if all lengths are scaled to the Hubble radius. Note, in particular, that this {\it scaling solution} is independent of the mass per unit length $\mu$ of the strings. 

The scaling solution is maintained by long strings losing energy into string loops with radii $R$ which are smaller than the Hubble scale.  We will adopt a {\it one-scale} model for the distribution of string loops \cite{onescale} according to which string loops are born at time $t$ with radius $R = \alpha t$, where $\alpha$ is a constant smaller than $1$. Loops are not exactly circular, and hence their length is larger than $2 \pi R$. We write the length as $l = \beta R$. Typical values of $\alpha$ and $\beta$ are \cite{CSsimuls} $\alpha = 0.1$ and $\beta = 10$ \footnote{See also \cite{AT} for original cosmic string evolution simulations.}.   The string distribution is also influenced by the mean number $N$ of infinite strings crossing any given Hubble volume. The numbers $N$, $\alpha$ and $\beta$ are in principle determined by physics (and independent of $\mu$).  It can be shown analytically \cite{CS-revs} that the network of infinite strings scales, but the exact parameters of the scaling solution must be determined by numerical simulations, and these simulations are challenging \cite{CSsimuls} because of the huge hierarchy of length scales between the cosmological length scale $t$ and the width $w$ of the strings which is proportional to $\eta^{-1}$. The simulations referred to above are based on an effective description of strings by the Nambu-Goto action (see \cite{CS-revs} for a discussion). In principle, full field theory simulations would be better. However, the problem of hierarchy of scales is more acute in such simulations (some of these field theory simulations in fact do not yield a scaling distribution of string loops \cite{Hind}). We will be assuming a scaling solution of string loops.
 
In the one-scale model for the string loop distribution, strings are created with a fixed radius $R = \alpha t$, and their number density then redshifts as space expands.  Due to gravitational radiation, the loop radius slowly shrinks \cite{GW}.  For loops with radius $R > \gamma \beta^{-1} G \mu t \equiv R_c(t)$ (where $\gamma$ is a constant of the order $10^2$ determined by the strength of gravitational radiation) the loop radius decay is negligible. However, loops with radius $R < \gamma \beta^{-1} G \mu t$ live less than one Hubble expansion time and have a negligible effect on structure formation. Based on these considerations, the number density in comoving coordinates $n_c(R, t) dR$ of loops in the radius interval between $R$ and $R + dR$ at time $t$ takes the form \cite{CS-revs}
\begin{widetext}
\be
n(R,t)=\left\{\begin{array}{ll}
N\alpha^2\beta^{-2}t_0^{-2}R^{-2}\ \ &\alpha t_{eq}<R\leq \alpha t\\
N\alpha^{5/2}\beta^{-5/2}t_{eq}^{1/2}t_0^{-2}R^{-5/2}\ \ &\gamma\beta^{-1}G\mu t_{eq}\leq R\leq \alpha t_{eq}\\
n(R_c(t),t)\ \ & R<\gamma\beta^{-1}G\mu t_{eq}
\end{array}\right. .\label{loopnumber1}
\ee
\end{widetext}
where $t_0$ is the present time and $t_{eq}$ is the time of equal matter and radiation.  The first line corresponds to loops formed in the matter era, the second and third lines are for loops formed during radiation domination. The number density of loops below the gravitational radiation cutoff is taken to be constant since all of these loops were formed within the same Hubble time (this is in fact an upper bound on the number density of these loops), but the exact form of the distribution for these loops will be irrelevant for our analysis.

%%%%%%%%%%%%%%%%%%%%%%%%%%%%%%%%%%%%%%%%%%%%%%%%%%%%%%%%%%%%%%%%%%%%%%%%%%%%%%%%%%%%%%%%

\section{Halo Mass Function Seeded by String Loops} \label{analysis}

Since string loops are localized overdensities, they will accrete matter. Because the mean separation of string loops is parametrically larger than the region over which a string loop accretes matter, it is a good approximation to consider that loops accrete independently from each other. 

At time $t$, a string loop of radius $R$ will have accreted a mass $M(R, t)$.  The mass function of loop-induced halos hence can easily be derived from the string loop distribution $n(R, t)$ from eq. (\ref{loopnumber1}):
\be
\frac{dn}{dM} \, = \, n(R(M),t)\frac{dR}{dM},
\ee
The relation between the loop-seeded non-linear halo mass and the loop radius is \cite{CS-revs}
\begin{widetext}
\be
M(R,z)=\left\{\begin{array}{ll}
\beta\mu R\left(\frac{z_f+1}{z+1}\right)\ \ &\alpha t_{eq}<R\leq \alpha t \\
\beta\mu R\left(\frac{z_{eq}+1}{z+1}\right)\ \ &\gamma\beta^{-1}G\mu t_{eq} \leq R \leq \alpha t_{eq} \\
\beta\mu \frac{R^2}{R_c(t_{eq})}\left(\frac{z_{eq}+1}{z+1}\right)\ & R<\gamma\beta^{-1}G\mu t_{eq}
\end{array}\right. .\label{loopnumber2}
\ee
\end{widetext}
where $z_f$ is the redshift when the loop was formed.  The first two lines represent the fact that the mass grows linearly as a function of inverse redshift between the time that the loop was formed (in the case of loops formed after $t_{eq}$) or $t_{eq}$ (in the case of loops formed before $t_{eq}$ but still present (and living for a Hubble time or more) at $t_{eq}$).  There are loops with $R < R_c(t_{eq})$ which are present at $t_{eq}$, but they have a highly suppressed mass accretion \cite{Pat} since they only live for a fraction $R / R_c(t_{eq})$ of a Hubble time.

To obtain the stellar mass function, we should consider stellar mass of loop-seeded galaxies instead of halos. Here we accept a simple relation between the stellar mass and halo mass:
\be
M_* \, = \, \epsilon f_b M,
\ee
where $f_b=0.156$ is the baryon fraction and $\epsilon$ is the star formation efficiency. This stellar mass comes from the physical picture of loop-seeded galaxies:
\begin{itemize}
\item[-] For loops formed in the radiation phase ($R\leq\alpha t_{eq}$), they begin to accrete dark matter at $t_{eq}$. These loop-seeded halos grow linearly \cite{CS-revs}. After recombination, baryons fall into the potential wells of dark matter halos and then form galaxies. Since the density distributions of both dark matter and baryons are almost uniform before accretion, we assume that the baryonic mass of a loop-seeded galaxy is roughly a fraction $f_b$ of the halo mass, i.e. $M_b=f_b M$ and the star formation efficiency $\epsilon$ denotes the fraction of baryons in stars, which is set to be 1 when evaluating the analytical equations.
\item[-] For loops generate after $t_{eq}$, they begin to accrete both dark matter and baryons immediately after their formation time $t_f$.  However, this does not affect the linear growth of galaxy and halo masses, as well as the ratio between them since we are interested in period before the sharp jump in regular galaxy formation.
\end{itemize}

For $\alpha t_{eq}<R\leq \alpha t$ (loops formed after $t_{eq}$) the stellar mass in the halo corresponding to the nonlinear density fluctuation seeded by the string loop is in the range $M_{eq}(z) < M_* < M_u(z)$ where
\begin{widetext}
\ba 
M_{eq}(z) \, &=& \, \epsilon f_b\alpha\beta\mu t_{eq}\frac{z_{eq}+1}{z+1}=7.0\times10^6M_\odot\cdot\epsilon\left(\frac{G\mu}{10^{-10}}\right)\left(\frac{z_{eq}+1}{z+1}\right) \ , \\
M_u(z) \, &=& \,  \epsilon f_b\alpha\beta \mu t=7.0\times10^6M_\odot\cdot\epsilon\left(\frac{G\mu}{10^{-10}}\right)\left(\frac{z_{eq}+1}{z+1}\right)^{3/2} \, .
\ea
\end{widetext}
The mass function corresponding to this range is
\begin{widetext}
\ba \label{mf1}
\frac{dn}{dM_*} \, &=& \, 3N\alpha^4\beta \epsilon^3 f_b^3 \mu^3 M_*^{-4}\left(z+1\right)^{-3}\\
&=& \, 9.9\times10^{22}M_\odot^{-1}Mpc^{-3}\cdot N\epsilon^3(1+z)^{-3}\left(\frac{G\mu}{10^{-10}}\right)^{3}\left(\frac{M_*}{M_\odot}\right)^{-4} \nonumber
\ea
\end{widetext}
Thus,  the comoving cumulative stellar mass density contained within galaxies above a certain stellar mass $M_*$ is
\begin{widetext}
\ba
\rho_*(>M_*,z) \, &=& \, \int_{M_*}^\infty \frac{dn}{dM_*}M_* dM_* \nonumber \\
&=& \, \frac32 N\alpha^4\beta \epsilon^3 f_b^3 \mu^3 M_*^{-2}\left(z+1\right)^{-3}\\
&=& \, 5.0\times10^{22}M_\odot Mpc^{-3}\cdot N\epsilon^3(1+z)^{-3}\left(\frac{G\mu}{10^{-10}}\right)^{3}\left(\frac{M_*}{M_\odot}\right)^{-2} \nonumber
\ea
\end{widetext}
 
For $\gamma\beta^{-1}G\mu t_{eq}\leq R\leq \alpha t_{eq}$ (loops formed before $t_{eq}$ but still present at $t_{eq}$ the stellar mass range is 
$M_c^{GW}(z) < M_* < M_{eq}(z)$ with
\ba
M_c^{GW}(z) \, &=& \, \epsilon f_b\beta\mu\left(\frac{z_{eq}+1}{z+1}\right)\cdot\gamma\beta^{-1}G\mu t_{eq} \\
&=& 0.07\,M_\odot\cdot \epsilon\left(\frac{G\mu}{10^{-10}}\right)^2\left(\frac{z_{eq}+1}{z+1}\right) \nonumber
%M_{eq}(z) \, &=& \, \epsilon f_b\alpha\beta\mu t_{eq}\frac{z_{eq}+1}{z+1}=7.0\times10^6M_\odot\cdot\epsilon\left(\frac{G\mu}{10^{-10}}\right)\left(\frac{z_{eq}+1}{z+1}\right)  \nonumber
\ea

Then, the stellar mass function is
\begin{widetext}
\ba \label{mf2}
\frac{dn}{dM_*} \, &=& \, N\alpha^{5/2}\beta^{-1}t_{eq}^{1/2}t^{-2}_0\left(\frac{z_{eq}+1}{z+1}\right)^{3/2}\epsilon^{3/2}f_b^{3/2}\mu^{3/2}M_*^{-5/2} \nonumber \\
&=& \, 2.9\times10^6 M_\odot^{-1}Mpc^{-3}\cdot N\epsilon^{3/2}(z+1)^{-3/2}\left(\frac{G\mu}{10^{-10}}\right)^{3/2}\left(\frac{M_*}{M_\odot}\right)^{-5/2}
\ea
\end{widetext}

Hence, the comoving cumulative stellar mass density contained within galaxies above a certain stellar mass $M_*$ is

\begin{widetext}
\ba
\rho_*(>M_*,z) \, &=& \, \int_{M_*}^\infty \frac{dn}{dM_*}M_* dM_* \nonumber \\
&=& \, 2N\alpha^{5/2}\beta^{-1}t_{eq}^{1/2}t^{-2}_0\left(\frac{z_{eq}+1}{z+1}\right)^{3/2}\epsilon^{3/2}f_b^{3/2}\mu^{3/2}M_*^{-1/2}\\
&=& \,  5.8\times10^{6} M_\odot Mpc^{-3}\cdot N\epsilon^{3/2}(z+1)^{-3/2}\left(\frac{G\mu}{10^{-10}}\right)^{3/2}\left(\frac{M_*}{M_\odot}\right)^{-1/2} \nonumber
\ea
\end{widetext}
 
For the sake of completeness, we can also consider loops with $R<\gamma\beta^{-1}G\mu t_{eq}$ (which have decayed before $t_{eq}$ or, if still present at $t_{eq}$, live for less than a Hubble time). In this case, the mass range is
%%
%\begin{widetext}
%%
\ba
M_* \, &<& \,  M_c^{GW}(z)= \epsilon f_b\gamma G\mu^2 t_{eq}\left(\frac{z_{eq}+1}{z+1}\right)
\ea
%%
%\end{widetext}
%%
Since these loops yield a negligible contribution to the cumulative mass function, we can take this mass function to be constant for $M_*<M_c^{GW}(z)$:
\be
\rho_*(>M_*,z) \, = \, \rho_*(>M_c^{GW}(z),z).
\ee

\begin{figure}
	\includegraphics[scale=0.5]{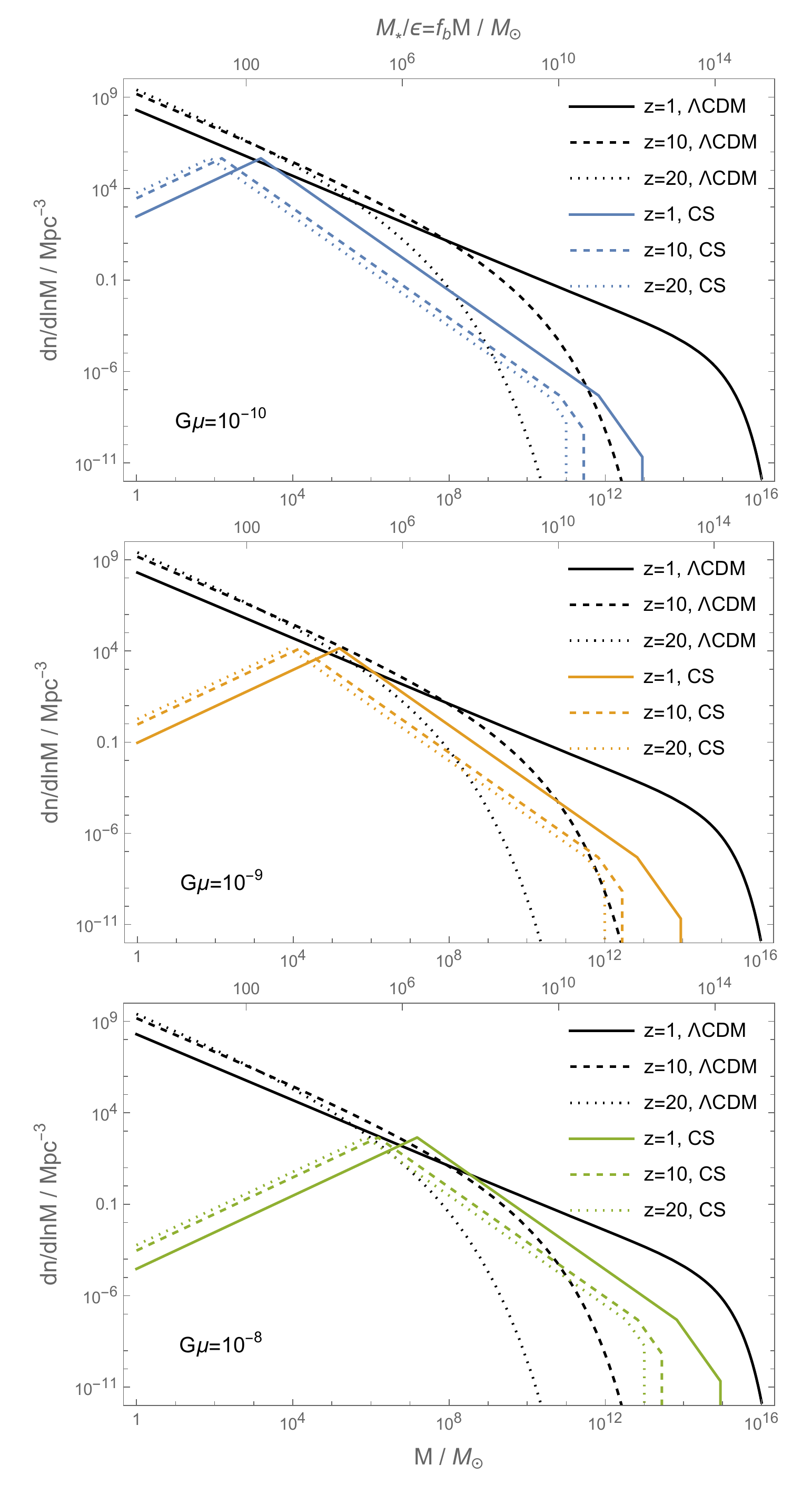}
	\caption{Comparison of the contribution of cosmic string loops to both the halo and stellar mass functions $dn/d\ln M$ with the predictions of the $\Lambda$CDM model for $G\mu=10^{-10},\ 10^{-9}$ and $10^{-8}$, from top to bottom, respectively. The values of the cosmic string parameters are given in the text.}
	\label{Fig1}
\end{figure}

If we set $f_b = 1$ and $\epsilon = 1$ in the above formulas, we obtain the halo mass function.  The first key lesson we can then derive from the above results is that the halo mass function decays only as a power of $(1 + z)$, not exponentially as the halo mass function in the $\Lambda$CDM model does. Hence, at high redshifts, the halo mass function is dominated by the contribution of cosmic string loops, even for values of $G\mu$ smaller than the upper bound from pulsar timing measurements.  The halo mass function from cosmic string loops is compared to that from the $\Lambda$CDM model for various redshifts in Fig. \ref{Fig1}. The three panels show (from top to bottom) the results for $G\mu = 10^{-10}, 10^{-9}$ and $10^{-8}$, respectively, at redshifts of $z = 10, 15$ and $20$.  The cosmic string parameters used were $\alpha = 0.1, \beta = 10, \gamma = 10^2$ and $N = 570$ which are the best-fit parameters from cosmic string simulations \cite{CSsimuls}.  The amplitude of the mass function depends on these parameters, but the slope does not.  The stellar mass functions can also be read off from the above results, modulated by a possible redshift dependence of the star formation efficiency $\epsilon$. Setting $\epsilon = 1$ we obtain stellar mass functions which are identical to the total halo mass function with a displaced mass axis.  In these plots, the horizontal axis is the halo mass in solar mass units (values below the graph) or stellar mass (values above the graph). The vertical axis is the mass function.

The graphs show that at low redshifts the $\Lambda$CDM fluctuations dominate the mass function.  On the other hand, at a redshift $z = 20$ the loop-seeded halos dominate the mass function, except for low masses which correspond to loops with radius below the gravitational radiation cutoff.  The transition redshift $z_c$ above which the contribution of strings dominates depends on the value of $G\mu$. For the values we studied here, the transition redshift occurs around $z = 10$. The transition redshift also depends mildly on the value of the mass which is being considered.

Note that the sharp upper cutoff in the mass stems from the fact that there is an upper cutoff on the loop radius.  This loop cutoff radius/mass is a function of both time and string tension $G\mu$. However, the induced upper cutoff in the mass function $\frac{dn}{d\ln M}$ is time-independent. There are two changes in the slope of the curves for the string-induced halo mass function. They are due to the change in the functional form of $M(R)$ at the value of $R$ corresponding to loops created at $t_{eq}$ (high mass transition point) and at value of $R$ corresponding to the gravitational radiation cutoff (low mass transition point). 

%%%%%%%%%%%%%%%

\begin{figure}[h!]
  \includegraphics[width=8cm]{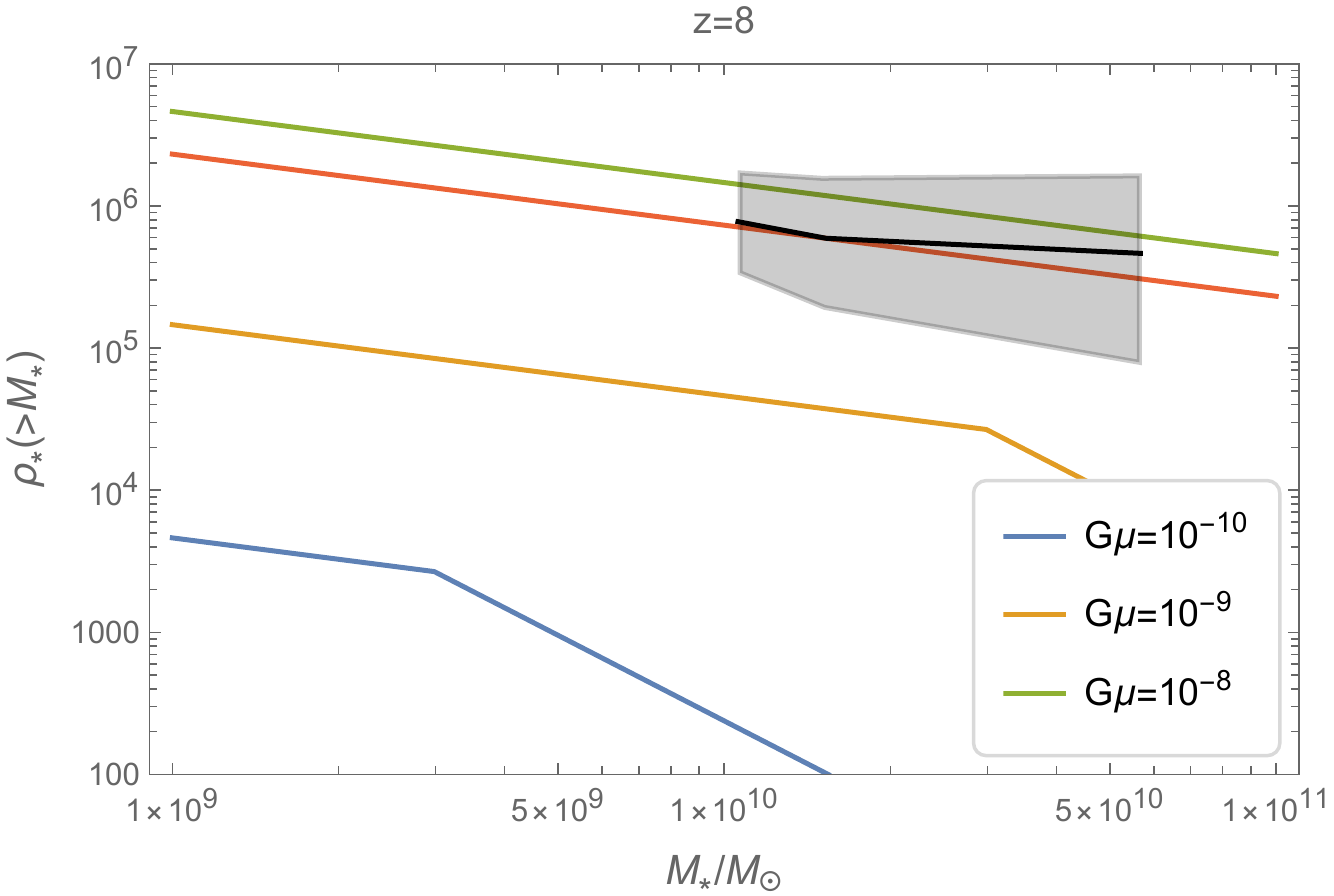}
\caption{Cumulative stellar mass density at redshift $z \simeq 8$: Predictions of the cosmic string model vs. preliminary JWST results. The four colorful lines are cumulative stellar mass densities of loop-seeded galaxies for $G\mu=10^{-10}$ (blue), $10^{-9}$ (orange), $10^{-8}$ (green), and $10^{-8.2}$ (red) separately. The black line with shaded region is the inferred mass density in the redshift bin $7 < z < 8.5$ from JWST data \cite{JWST-heaviest}. We are assuming a simple linear scaling between halo mass and stellar mass with star formation efficiency $\epsilon = 1$.}
\label{Fig2}
\end{figure}

\begin{figure}[h!]
  \includegraphics[width=8cm]{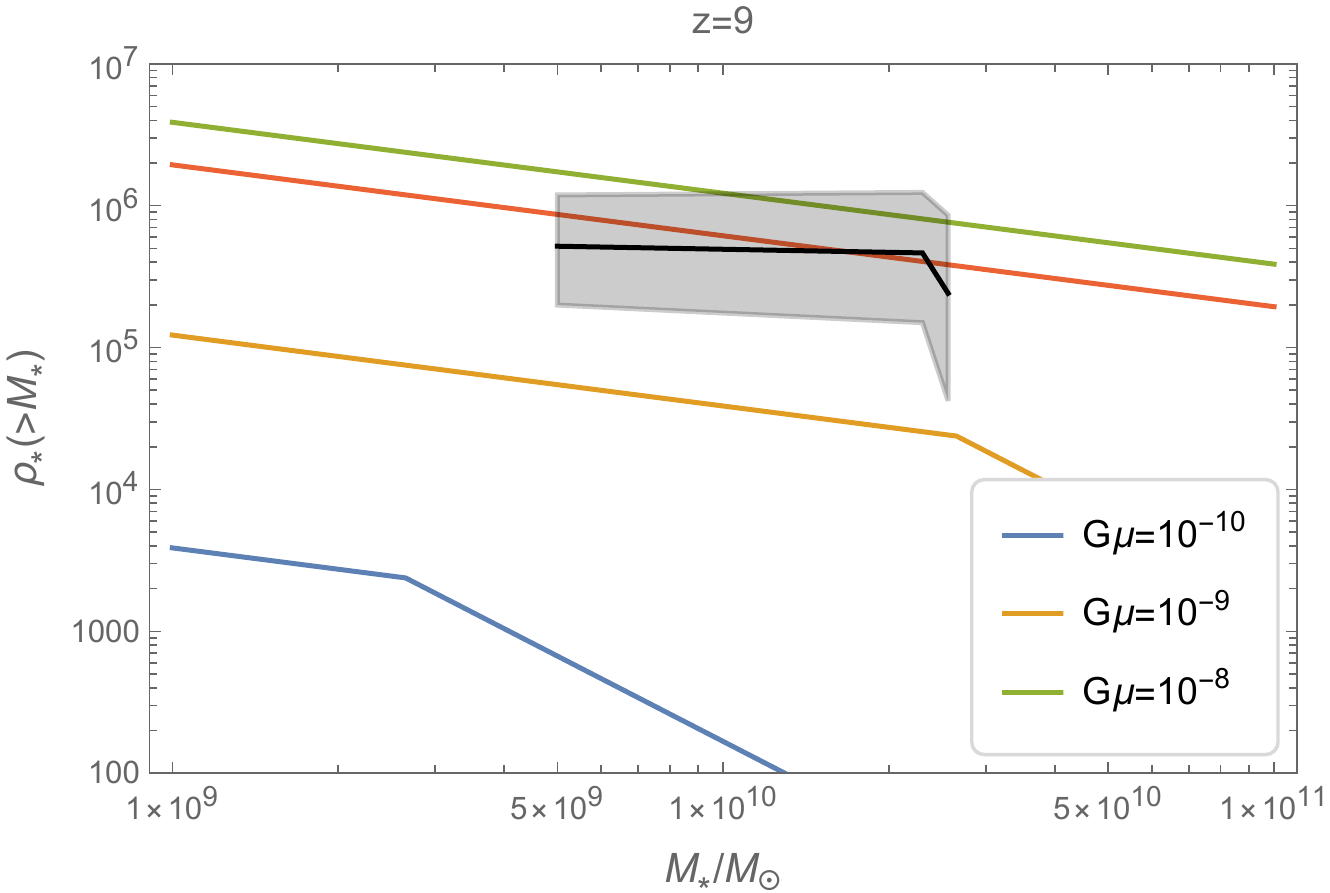}
\caption{Cumulative stellar mass density at redshift $z \simeq 9$: Predictions of the cosmic string model vs. preliminary JWST results \cite{JWST-heaviest}.}
\label{Fig3}
\end{figure}

\begin{figure}[h!]
  \includegraphics[width=8cm]{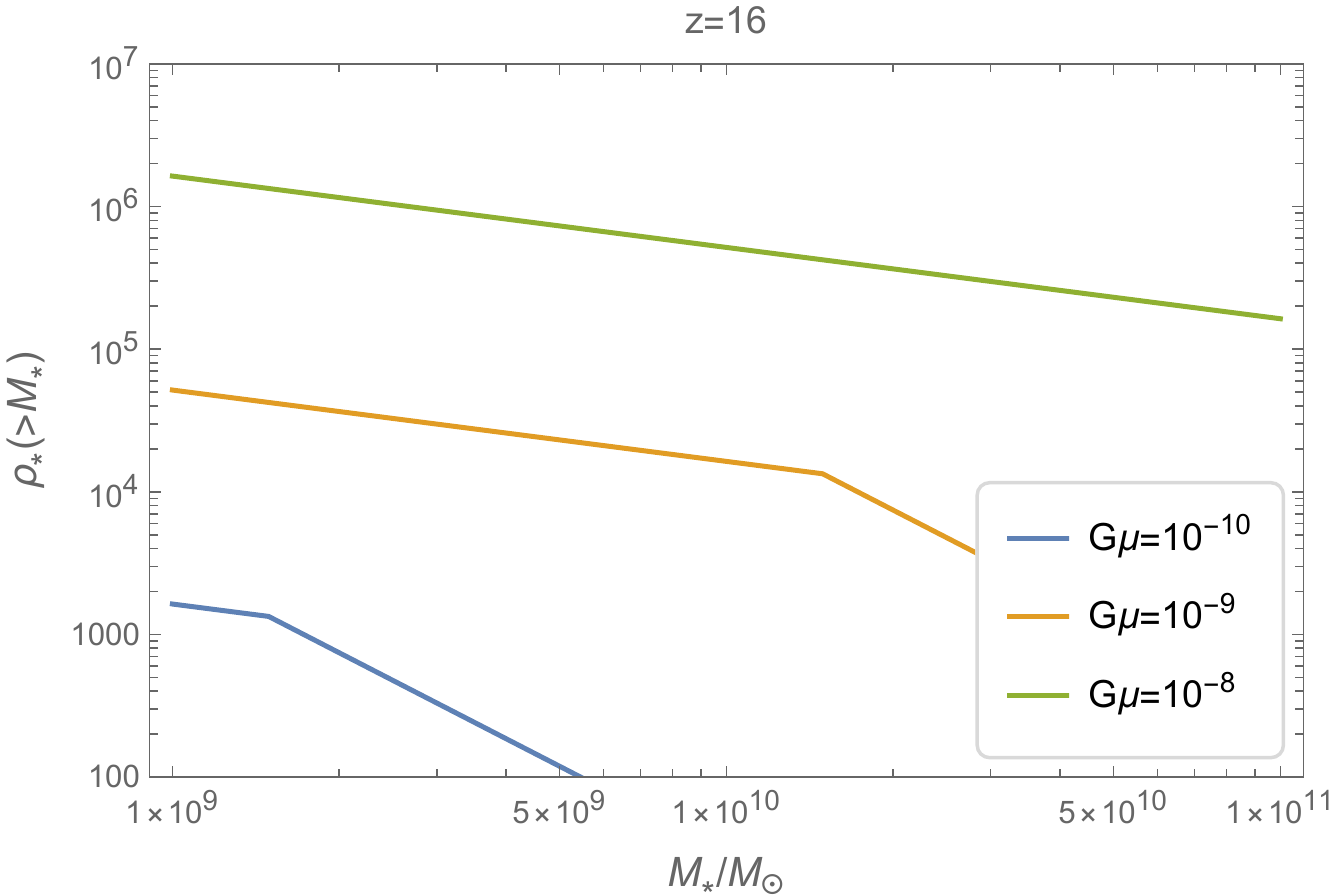}
\caption{Predicted cumulative stellar mass density at redshift $z = 16$.}
\label{Fig4}
\end{figure}

It is the stellar mass function which is relevant if we want to compare our predictions to the recent JWST results. In Figs. \ref{Fig2} and \ref{Fig3} we compare the comoving cumulative stellar mass density contained within galaxies above a certain stellar mass $M_*$. With a value of $G\mu \sim 10^{-8.2}=6.3\times10^{-9}$, the current JWST data at redshifts $z = 8$ and $z = 9$ \cite{JWST-heaviest} can both be well matched.  Assuming that string loops provide an explanation for the current data,  the string model then makes specific predictions for the shape of the mass function and its redshift dependence which JWST and other experiments should be measuring once more data comes in.  For example,  the prediction of the comoving cumulative stellar mass density contained within galaxies above a certain stellar mass $M_*$ at redshift $z = 16$ is shown in Fig. \ref{Fig4} (this redshift was chosen since preliminary JWST data indicate the existence of galaxies at this redshift \cite{early}.) If string loops contribute only a fraction $F < 1$ of the halo mass function inferred from the JWST data, the string model predicts the amplitude and shape of the halo mass function at higher redshifts, and for redshifts higher than some critical value $z_c(F)$ the contribution of string loops would dominate. 

\begin{figure}[h!]
  \includegraphics[width=8cm]{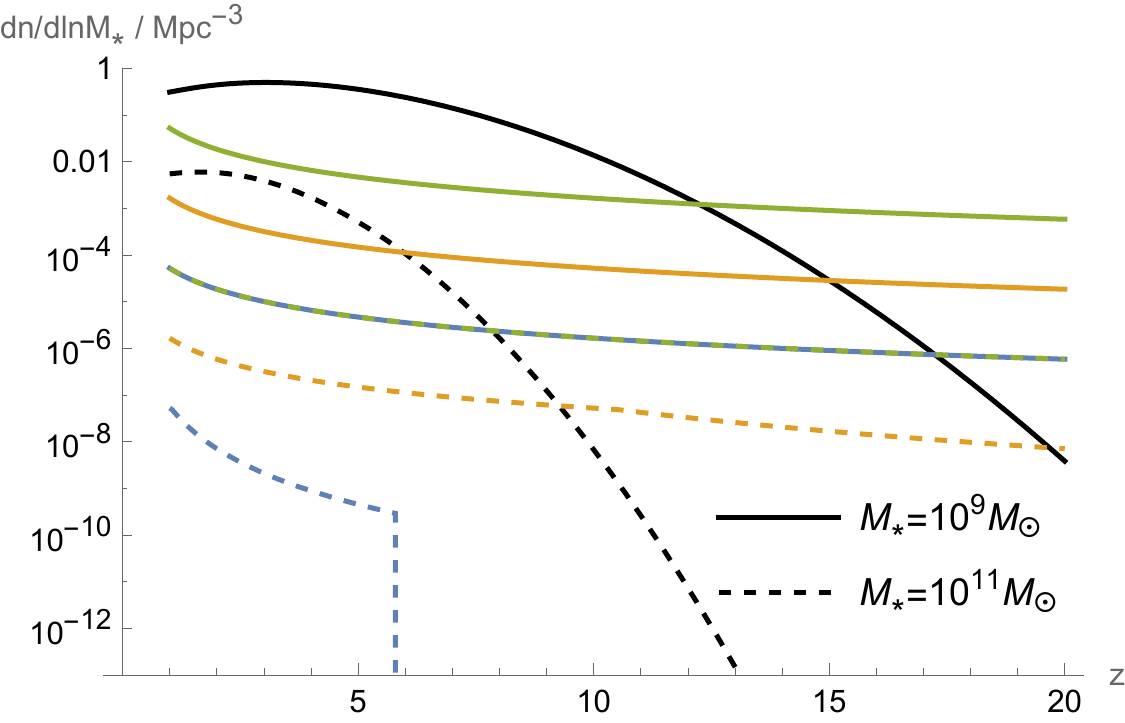}
\caption{Stellar mass function as a function of redshift for two representative values of the mass related to the high-redshift galaxy candidates detected by JWST: Predictions of the cosmic string vs. $\Lambda$CDM model. The black lines are the predictions of the $\Lambda$CDM model, and the colored curves are the predictions of the string model, with $G\mu = 10^{-10}$ (blue),  $G\mu = 10^{-9}$ (orange) and $G\mu = 10^{-8}$ (green). Note the accidental overlap between the blue solid curve and the green dashed line. }
\label{Fig5}
\end{figure}

In Fig. \ref{Fig5} we plot the stellar mass function $\frac{dn}{d\ln M}$ as a function of redshift for two representative values of the mass $M_*$. The solid lines correspond to $M_* = 10^9 \Msol$ and the dashed curves to $M_* = 10^{11} \Msol$. The black curves are for the $\Lambda$CDM model and the colored curves represent the contribution of cosmic string loops with the same color coding as in the previous figures, i.e.  for $G\mu = 10^{-10}$ in blue, for $G\mu = 10^{-9}$ in orange, and for $G\mu = 10^{-8}$ in green. Note that the solid blue curve overlaps with the dashed green curve.  This figure represents another demonstration of the fact that the contribution of cosmic strings will dominate at high redshifts as long as the mass is lower than the mass of the largest loop at time $t$.

%%%%%%%%%%%%%%%%%%%%%%%%%%%%%%%%%%%%%%%%%%%%%%%%%%%%%%%%%%%%%%%%%%%%%%%%%%%%%%%%%%%%%%%%%%%%%%
 
\section{Extreme Value Statistic for the Expected Mass of the Heaviest Halo}
 
The {\it extreme value statistics} (EVS) method \cite{JWST-EVS, EVS} is a good way to compare a mass function with an extreme observation.

Given a survey volume of the sky,  and assuming that there are $N_{tot}$ halos which are detected,  then the probability $\Phi(M_{max}\leq m;N_{tot})$ that all halos have a mass $M$ less than or equal to some value $m$ is given by the product
\ba \label{eq-CDFPhi}
\Phi(M_{max}\leq m;N_{tot}) \, &=& \, F_1(M_1\leq m)\cdots F_{N_{tot}}(M_N\leq m) \nonumber \\
&=& \, F^{N_{tot}}(M)
\ea
where in the second step we have assumed that the halos are drawn from a single mass distribution function $F(M \leq m)$,  i.e. $F_i(M \leq m) = F(M \leq m)$ for all $i$ (where $F_i(M \leq m)$ is the probability that the mass $M$ of the i'th halo is smaller or equal to $m$).

The probability density function (PDF) of the above distribution is
\ba \label{eq-PDFPhi}
\Phi(M_{max} = m_*;N_{tot}) \, &=& \, N_{tot}F'(m_*)F^{N_{tot}-1}(m_*)  \nonumber \\
&\equiv& \, N_{tot}f(m_*)F^{N_{tot}-1}(m_*) 
\ea
where $\Phi(M_{max} = m_*;N_{tot}) dm_*$ is the probability that the maximum mass for a sample of $N$ halos is in the range between $m_*$ and $m_* + dm_*$. Note that the distribution function $f(m_*)$ is defined by the above cumulative distribution function $F(M)$.

Assuming that one is observing halos in the redshift interval between $z_{min}$ and $z_{max}$, then the relation between the two distribution functions and the galaxy mass function is
\ba \label{eq-PDFf}
f(m_*) \, &=& \, \frac1{N_{tot}} \int_{z_{min}}^{z_{max}}dz\frac{dV_c}{dz}\frac{dn(m_*,z)}{dm_*}\\ \label{eq-CDFF}
F(m_*) \, &=& \,\int_{0}^{m_*}dM_*f(M_*)\\
&=& \, \frac1{N_{tot}}\int_{0}^{m_*}dM_* \int_{z_{min}}^{z_{max}}dz\frac{dV_c}{dz}\frac{dn(m_*,z)}{dm_*} \nonumber
\ea
where $V_c(z)$ is the comoving Hubble volume at redshift $z$ corresponding to the survey sky area. 

In the following we will compute the probability distribution of the maximal stellar mass in the distribution of halos to be observed in the region of sky and redshift coverage of the released JWST telescope data, assuming that the halos are seeded by a scaling distribution of cosmic string loops.

The starting point is the mass function of (\ref{mf1}) and (\ref{mf2}) which follows immediately from the loop scaling distribution (\ref{loopnumber1}) (since halos created by loops which have already decayed bt $t_{eq}$ are negligible, we can set the corresponding number density to zero).

\begin{figure}[h!]
  \includegraphics[width=8cm]{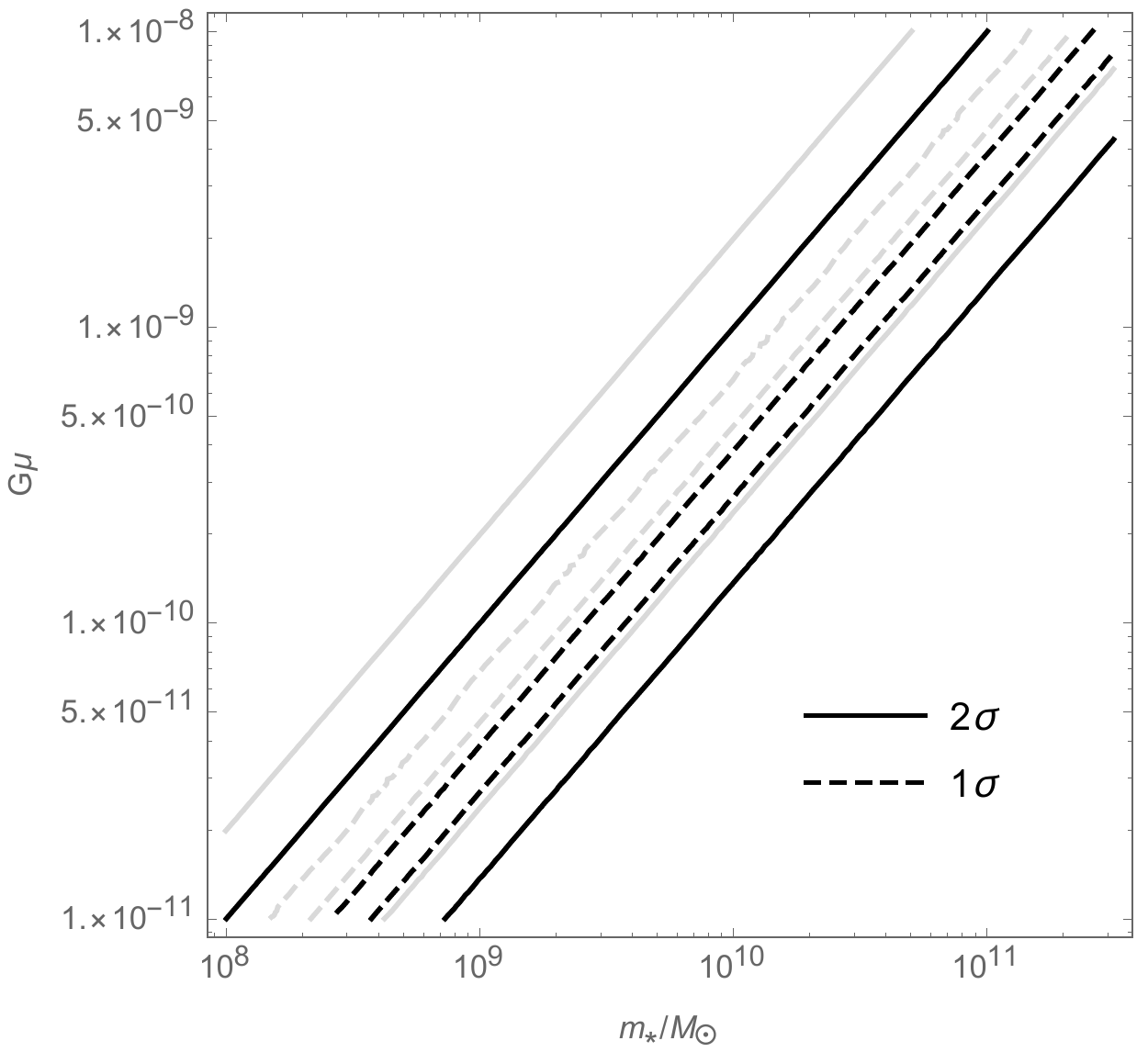}
\caption{The 1$\sigma$ (dashed) and 2$\sigma$(solid) confidence intervals of the EVS PDF of the expected largest solar mass halo predicted by the cosmic string model as a function of $G\mu$ evaluated for the angular and redshift coverage corresponding to the preliminary JWST data. The horizontal axis is the mass, the vertical axis is $G\mu$. The black lines correspond to redshift $z=10$ and the gray lines are $z=16$.} %% Since we talked about the standard value of N in the previous text, I didn't mention this in this caption.
\label{Fig6}
\end{figure}

\begin{figure}[h!]
  \includegraphics[width=8cm]{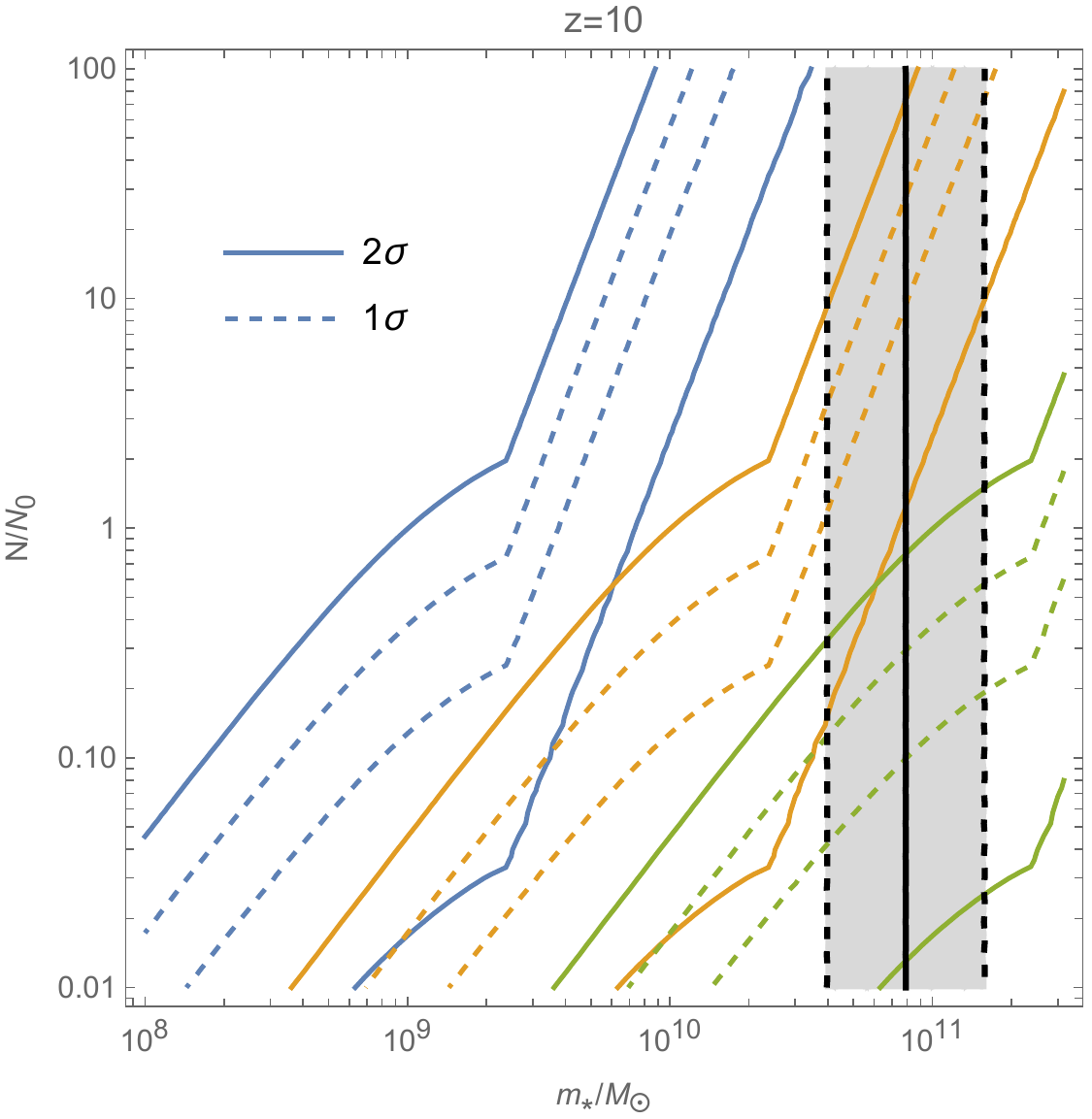}
\caption{Expected largest solar mass halo for a survey with the sky and redshift coverage of the preliminary JWST analyses.  The vertical axis represents the cosmic string distribution parameter $N$ plotted in units of its canonical value $N_0=570$, and the horizontal axis is the stellar mass.  One and two sigma confidence intervals are shown. The colors correspond to the same three different values of $G\mu$ used earlier (see text).  The predictions are for redshift $z \geq 10$. The vertical black lines with gray shaded region correspond to the inferred stellar mass of Galaxy 14924, which is the heaviest high-redshift galaxy from JWST.}
\label{Fig7}
\end{figure}

\begin{figure}[h!]
  \includegraphics[width=8cm]{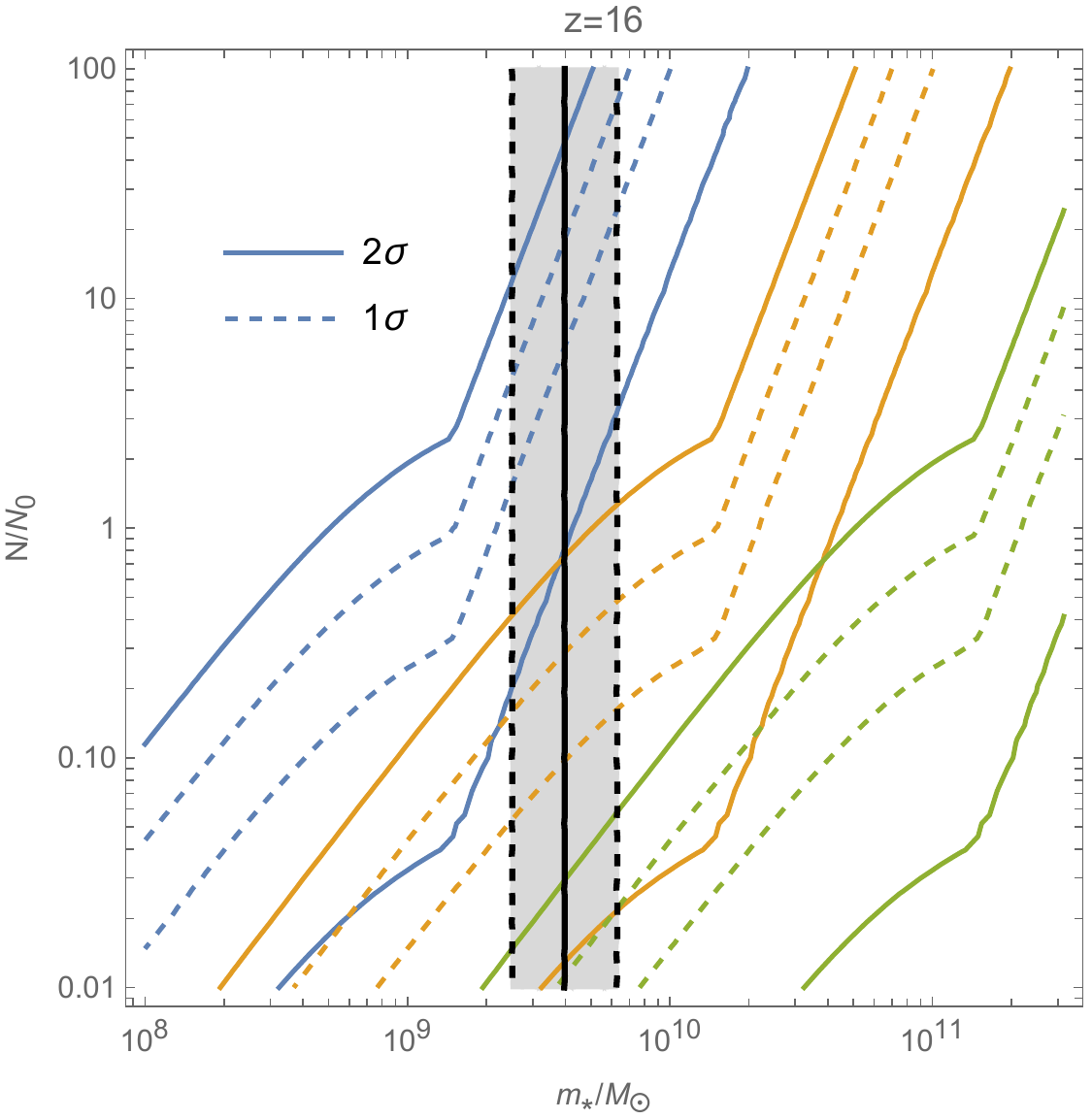}
\caption{The same as in Fig. (\ref{Fig7}),  but for redshift $z \geq 16$ The shaded region corresponds to the JWST galaxy CEERS-1749.}
\label{Fig8}
\end{figure}

The analytical computation of the PDF $\Phi(m)$ is summarized in the Appendix. % There, it is shown that the contribution of loops formed in the matter epoch to the probability distribution of the heaviest stellar mass is
%%
%\begin{widetext}
%%
%\ba
%\Phi(M_{max}=m_*) \, &=& \, N_{tot}f(m_*)F^{N_{tot}-1}(m_*) \, \simeq \, N_{tot}f(m_*)\\
%&=& \, 4\pi N\alpha^{5/2}\beta^{-1}\epsilon^{3/2}f_b^{3/2}f_{sky}H_0^{-3}\Omega_M^{-3/2}G^{-3/2} t_{eq}^{-1/2}t^{-1}_0(G\mu)^{3/2} m_*^{-5/2}(1+z_{min})^{-2} \nonumber
%\ea
%%
%\end{widetext}
%%
%while the contribution of loops formed in the radiation epoch is
%%
%\begin{widetext}
%%
%\ba
%\Phi(M_{max}=m_*) \, &=& \, N_{tot}f(m_*)F^{N_{tot}-1}(m_*) \, \simeq \, N_{tot}f(m_*)\\
%&=& \, 4\pi N\alpha^{5/2}\beta^{-1}\epsilon^{3/2}f_b^{3/2}f_{sky}H_0^{-3}\Omega_M^{-3/2}G^{-3/2} t_{eq}^{-1/2}t^{-1}_0(G\mu)^{3/2} m_*^{-5/2}(1+z_{min})^{-2} \, . \nonumber
%\ea
%%
%\end{widetext}
%%
%
The results of the EVS statistic are represented in Figs. \ref{Fig6} - \ref{Fig8}. Figure \ref{Fig6} shows the confidence intervals of the EVS PDF of the largest solar mass halo (horizontal axis) as a function of $G\mu$ (vertical axis). The black and gray lines correspond to redshifts $z \geq 10$ and $z \geq 16$ separately. This extreme value statistic is evaluated for the sky and redshift coverage of the preliminary JWST analyses, with a cosmic string scaling solution with the standard parameter value $N = 576$. % We see that for $G\mu > 3 \times 10^{-10}$ it is possible to explain the origin of halos with stellar mass of about $10^{10} \Msol$.

The EVS statistic is sensitive to the total number of objects in the sample. Hence, the largest expected halo mass will be sensitive to the string distribution parameter $N$.  In Figs. \ref{Fig7} and \ref{Fig8} we show how the predicted mass of the largest halo (again for the sky and redshift coverage corresponding to the preliminary JWST analyses) depends on the string distribution parameter $N$. The vertical axis is the parameter $N$ in units of its canonical value $N_0=570$, and the horizontal axis gives the mass. The color coding corresponds to the same values of $G\mu$ used earlier: $G\mu = 10^{-8}$ in green, $G\mu = 10^{-9}$ in orange, and $G\mu = 10^{-10}$ in blue. One and two $\sigma$ error bars are indicated. Note that the kinks in the curves are due to the transition between loops created in the matter and radiation epoch dominating.

%% The paragraph I added
We compare the predicted stellar mass of the largest loop-seeded galaxy to the two extreme galaxies detected by JWST: 
\begin{itemize}
\item[-] The heaviest hight-redshift galaxy candidate: Galaxy 14924, which has stellar mass $\log_{10}(M_*/M_{\odot})=10.9 \pm 0.3$ and inferred redshift $ z = 9.9 \pm 0.5$ \cite{JWST-heaviest}.
\item[-] The earliest galaxy candidate: CEERS-1749 with stellar mass $\log_{10}(M_*/M_\odot)= 9.6\pm0.2$ and redshift $z = 16 \pm 0.6$ \footnote{There are two galaxy candidates at $z\sim 16$. They have similar inferred redshift but this one is heavier so we consider CEERS-1749 here.} \cite{JWST-data}.
\end{itemize}
We illustrate the $1\sigma$ range of stellar masses of the two extreme galaxies by the gray region with black vertical lines. We can find that for $N=N_0$, cosmic string loops with $G\mu\sim 10^{-9}$ are able to seed enough high-redshift halos to explain the extreme JWST galaxies. Besides, loop-seeded halos have more advantages for higher redshift cases and we only need $G\mu\sim 10^{-10}$ to generate CEERS-1749 at redshift $z\gsim 16$.

%%%%%%%%%%%%%%%%%%%%%%%%%%%%%%%%%%%%%%%%%%%%%%%%%%%%%%%%%%%%%%%%%%%%%%%%%%%%%%%%%%%%%%%%%%%%%%
\section{Conclusions and Discussion} \label{conclusion}

Assuming a simple linear relation between stellar and halo mass, we have studied the halo and stellar mass functions induced by a scaling distribution of cosmic strings, and compared the results with the predictions of the standard Gaussian $\Lambda$CDM model.  As to be expected, the contribution of the cosmic strings dominates at high redshifts since the halo mass function induced by cosmic strings decreases only as a power of $(1 + z)^{-1}$ rather than exponentially. The critical redshift above which the cosmic string contribution dominates depends on the value of the string tension, $G\mu$, and on the string distribution parameters $N, \alpha$ and $\beta$ which in principle are determined by the physics but not yet known exactly due to uncertainties in the numerical simulations of cosmic string dynamics.

We find that for a value of $G\mu \sim 10^{-8}$, the stellar halo mass densities inferred by the preliminary JWST data at both redshifts $z = 8$ and $z = 10$ can be matched by loop-seeded halos. With this normalization of the cosmic string model, the model makes specific predictions for what upcoming high redshift observations should detect (for example for the mass functions at redshift $z = 16$ which will be determined in upcoming JWST analyses).

An extreme value statistic analysis shows that we can explain the heaviest and earliest high-redshift galaxy candidates detected by JWST given a string tension as low as $G\mu\sim 10^{-9}$. Specifically, we use EVS to compute the stellar mass distribution of the heaviest loop-seeded halos for redshift $z>10$ and $z>16$, which correspond to the inferred redshift of the two extreme JWST galaxies.

The value $G\mu = 10^{-8}-10^{-9}$ is well below the current robust limit on the cosmic string tension coming from CMB anisotropy measurement, but it is larger than the limit inferred from pulsar timing array studies. Those limits, however, depend crucially on the full string loop distribution, in particular on the effective low radius cutoff. Increasing this cutoff would render the pulsar timing constraints consistent with the best-fit value of $G\mu$.

In our current analysis, we have treated the cosmic string loop as a point mass.  For values of $G\mu$ lower than the ones we are considering here, the effects of the finite extent of the string source in the gravitational accretion process become important. This will be discussed in a forthcoming publication \cite{JH}.

%%%%%%%%%%%%%%%%%%%%%%%%%%%%%%%%%%%%%%%%%%%%%%%%%%%%%%%%%%%%%%%%%%%%%%%%%%%%%%%%%%%%%%%%%%%%%%
\section*{Acknowledgement}

\noindent RB wishes to thank the Pauli Center and the Institutes of Theoretical Physics and of Particle- and Astrophysics of the ETH for hospitality. The research of RB at McGill is supported in part by funds from NSERC and from the Canada Research Chair program.  

%%%%%%%%%%%%%%%%%%%%%%%%%%%%%%%%%%%%%%%%%%%%%%%%%%%%%%%%%%%%%%%%%%%%%%%%%%%%%%%%%%%%%%%%%%%%%%
\section*{Appendix}

In this appendix, we analytically derive the PDF of the heaviest loop-seeded galaxies
\be
\Phi(M_{*,max}=m_*) \, = \, N_{tot}f(m_*)F^{N_{tot}-1}(m_*)
\ee
by the EVS.

First, we need to calculate the total number of loop-seeded galaxies in the sky region observed by JWST:
\be
N_{tot} \, = \, \int_{z_{min}}^{z_{max}}dz\frac{dV_c}{dz}\int_{0}^{+\infty}dM_*\frac{dn(M_*,z)}{dM_*}.
\ee
where $\frac{dn(M_*,z)}{dM_*}$ is the mass function of loop-seeded galaxies in Eq. (\ref{eq-dndM}), and $V_c$ is the comoving Hubble volume at redshift $z$ and $\frac{dV_c}{dz}$ is:
\be
\frac{dV_c}{dz} \, = \, 4\pi f_{sky}D_H\frac{(1+z)^2 D_A^2}{E(z)},
\ee
where $f_{sky}=2.7\times 10^{-7}$ is the fraction of sky of the released JWST telescope data, $D_H$ is the Hubble distance
\be
D_H \, \equiv \, \frac{1}{H_0} \, = \, 9.26\times 10^{25} h^{-1} m,
\ee
and $D_A$ is the angular diameter distance
\ba
D_A\, &=& \, \frac{D_M}{1+z} \, = \, \frac{D_H}{1+z}\int_0^{z}\frac{dz'}{E(z')},\\
E(z)\, &\equiv& \, \sqrt{\Omega_M(1+z)^3+\Omega_k(1+z)^2+\Omega_\Lambda},
\ea
where, as usual, $\Omega_m$, $\Omega_k$ and $\Omega_\Lambda$ are the fractional energy densities in matter, spatial curvature, and dark energy, respectively. Since we are interested in high-redshift galaxies with $z \geq 10$, it is a good approximation to use $$E(z)\sim \sqrt{\Omega_M(1+z)^3}.$$ 

With this approximation, we can calculate the derivative:
\ba
\frac{dV_c}{dz} \, &=& \, 4\pi f_{sky}D_H^3\frac{1}{E(z)}\left(\int_0^{z}\frac{dz'}{E(z')}\right)^2 \nonumber \\
&\simeq & \, 8\pi f_{sky}H_0^{-3}\Omega_m^{-3/2}(1+z)^{-3/2}
\ea
Then, we can calculate the total galaxy number. Note that we aim to compare the PDF of the extreme galaxies with redshift greater than a certain value to JWST candidates, so we set $z_{max}=\infty$ and $z_{min}=z_{JWST}$ here.
\begin{widetext}
\ba
N_{tot} \, &=& \, \int_{z_{JWST}}^{\infty}dz\frac{dV_c}{dz}\int_{0}^{+\infty}dM_*\frac{dn(M_*,z)}{dM_*} \nonumber \\
&\simeq& \, \frac{2}{3} N\alpha^{5/2}\beta^{-1}\gamma^{-3/2}t_{eq}^{-1}t_0^{-2} (G\mu)^{-3/2} \int_{z_{JWST}}^{\infty}dz\frac{dV_c}{dz} \nonumber \\
&=& \, \frac{32\pi}{3} N\alpha^{5/2}\beta^{-1}\gamma^{-3/2} f_{sky}H^{-3}\Omega_m^{-3} t_{eq}^{-1}t_0^{-2} (G\mu)^{-3/2} (1+z_{JWST})^{-1/2} \nonumber \\
&=& \, 3.1\times10^{9}\cdot N\left(\frac{f_{sky}}{2.7\times10^{-7}}\right)\left(\frac{G\mu}{10^{-10}}\right)^{-3/2}(1+z_{JWST})^{-1/2},
\ea
\end{widetext}
where we have inserted the values $h=0.7$ and $\Omega_m=0.32$.

The galaxy distribution $f(m_*)$ due to cosmic string loops formed in the matter and radiation epochs ($M_*\geq M_{eq}(z)$ or $M_*<M_{eq}(z)$) should be calculated separately. For loops formed after $t_{eq}$, the PDF and CDF of the galaxy distribution are
\begin{widetext}
\ba
f(m_*) \, &=& \, \frac1{N_{tot}} \int_{z_{JWST}}^{\infty}dz\frac{dV_c}{dz}\frac{dn(m_*,z)}{dm_*} \nonumber \\
&=& \, \frac{9}{14}\alpha^{3/2}\beta^{2}\gamma^{3/2}\epsilon^3 f_b^3 G^{-3}t_{eq}t_0^{2} \left(G\mu\right)^{9/2} m_*^{-4}(1+z_{JWST})^{-3} \nonumber \\
&=& \, 2.4\times10^{-38}GeV^{-1}\cdot \epsilon^3 \left(\frac{G\mu}{10^{-10}}\right)^{9/2} \left(\frac{m_*}{M_\odot}\right)^{-4}(1+z_{JWST})^{-3}, \label{eq-PDFfMD}\\
F(m_*)&=& \, 1-\int_{m_*}^\infty dM_* f(M_*)\nonumber \\
&=& \, 1-\frac{6}{7}\alpha^{3/2}\beta^{2}\gamma^{3/2}\epsilon^3 f_b^3 G^{-3}t_{eq}t_0^{2} \left(G\mu\right)^{9/2}(1+z_{JWST})^{-3} m_*^{-3}\nonumber \\
&=& \, 1-3.6\times10^{19}\times \epsilon^3\left(\frac{G\mu}{10^{-10}}\right)^{9/2}\left(\frac{m_*}{M_\odot}\right)^{-3}(1+z_{JWST})^{-3}.\label{eq-PDFFMD}
\ea
\end{widetext}

For loops formed in the radiation-dominated era, i.e.  for galaxy masses $M_*<M_{eq}(z)$,
\begin{widetext}
\ba
f(m_*) \, &=& \, \frac1{N_{tot}} \int_{z_{JWST}}^{\infty}dz\frac{dV_c}{dz}\frac{dn(m_*,z)}{dm_*}\nn
&=& \, \frac{3}{8}\gamma^{3/2}\epsilon^{3/2}f_b^{3/2}G^{-3/2}t_{eq}^{1/2}t_0 (G\mu)^{3}m_*^{-5/2}(1+z_{JWST})^{-3/2}\nn
&=& \, 1.2\times10^{-54}GeV^{-1}\cdot\epsilon^{3/2}\left(\frac{G\mu}{10^{-10}}\right)^3\left(\frac{m_*}{M_\odot}\right)^{-5/2}(1+z_{JWST})^{-3/2}\label{eq-PDFfRD}\\
F(m_*)\, &=& \, 1-\int_{m_*}^{M_{eq}} dM_* f(M_*)-\int_{M_{eq}}^\infty dM_* f(M_*).\label{eq-PDFFRD}
\ea
\end{widetext}

From the above equations, we can get the distribution of the heaviest galaxy for redshift greater than $z_{JWST}$ by inserting eq. (\ref{eq-PDFfMD}--\ref{eq-PDFFRD}) into eq. (\ref{eq-CDFPhi}) and (\ref{eq-PDFPhi}). Here we use the logarithmic form to reduce the influence of rounding-off errors:
\ba
\ln \Phi(M_{*,max}\leq m_*) \, &=& \, N_{tot}\ln[F(m_*)]\\
\ln \Phi(M_{*,max}=m_*) \, &=& \, \ln\left[N_{tot}\right]+\ln\left[f(m_*)\right]\nn
&+& \, (N_{tot}-1)\ln\left[F(m_*)\right].
\ea

\end{document}